\documentclass[aps,prd,twocolumn,nofootinbib,preprintnumbers,bibnotes,superscriptaddress]{revtex4-2}

\usepackage{style}

\begin{document}

\title{MOSAIC: Magnonic Observations of Spin-dependent Axion-like InteraCtions}

\author{Clarence Chang\,\orcidlink{0000-0002-6311-0448}}
\email{clchang@anl.gov}
\affiliation{High Energy Physics, Argonne National Laboratory, Lemont, Illinois 60439, USA}
\affiliation{University of Chicago, Chicago, Illinois 60637, USA}
\affiliation{Kavli Institute for Cosmological Physics, University of Chicago, Chicago, Illinois 60637, USA}

\author{T. J. Hobbs\,\orcidlink{0000-0002-2729-0015}}
\affiliation{High Energy Physics, Argonne National Laboratory, Lemont, Illinois 60439, USA}

\author{Dafei Jin\,\orcidlink{0000-0002-8035-3580}}
\affiliation{Department of Physics and Astronomy, University of Notre Dame, Notre Dame, Indiana 46556, USA}

\author{Yi Li\,\orcidlink{0000-0002-3207-7147}}
\affiliation{Materials Science Division, Argonne National Laboratory, Lemont, Illinois 60439, USA}

\author{Marharyta Lisovenko\,\orcidlink{0000-0002-2946-3944}}
\affiliation{High Energy Physics, Argonne National Laboratory, Lemont, Illinois 60439, USA}

\author{Valentine Novosad\,\orcidlink{0000-0003-2127-2374}}
\affiliation{Materials Science Division, Argonne National Laboratory, Lemont, Illinois 60439, USA}

\author{Zain H. Saleem\,\orcidlink{0000-0002-8182-2764}}
\affiliation{Mathematics and Computer Science Division, Argonne National Laboratory, Lemont, Illinois 60439, USA}

\author{Tanner Trickle\,\orcidlink{0000-0003-1371-4988}}
\email{ttrickle@fnal.gov}
\affiliation{Fermi National Accelerator Laboratory, Batavia, Illinois 60510, USA}

\author{Gensheng Wang\,\orcidlink{0000-0003-1385-3806}}
\affiliation{High Energy Physics, Argonne National Laboratory, Lemont, Illinois 60439, USA}

\begin{abstract}

We introduce an array-scalable, magnon-based detector (MOSAIC) to search for the spin-dependent interactions of electron-coupled axion dark matter. These axions can excite single magnons in magnetic targets, such as the yttrium iron garnet (YIG) spheres used here, which are subsequently sensed by the detector. For MOSAIC, this sensing is implemented by coupling the magnons in the YIG spheres to magnetic-field-resilient single-electron charge-qubits, whose state is then interrogated with a quantum non-demolition measurement. Using standard superconducting fabrication techniques, MOSAIC can integrate many YIG sphere-qubit sensors, forming a large detector array. We outline the detector design and operation, and determine its sensitivity to axion dark matter. We find that a detector built with available technology will exceed the sensitivity of previous ferromagnetic haloscopes, and provides a platform where further improvements in performance would search for electron-coupled axion dark matter in unexplored parameter space.

\end{abstract}
\maketitle

An abundance of astrophysical and cosmological evidence points to the existence of dark matter (DM). One compelling class of DM candidates are light bosons, with mass less than an eV~\cite{Antypas:2022asj,Berlin:2024pzi}. The most theoretically motivated light boson is the QCD axion, a pseudo-Nambu-Goldstone boson generated from the breaking of an axial $U(1)_\text{PQ}$ Peccei-Quinn symmetry, since it can alleviate the strong CP problem~\cite{Peccei:1977ur,Peccei:1977hh,Wilczek:1977pj,Weinberg:1977ma}. More broadly, light bosons appear in string theory from compactified dimensions~\cite{Arvanitaki:2009fg,Svrcek:2006yi,Cicoli:2012sz} or from breaking of a high-scale global $U(1)$ symmetry. They do not have to solve the strong CP problem, and can have a variety of couplings with the Standard Model. Such pseudoscalars are sometimes referred to as ``axion-like" particles; we refer to them here as just ``axions" for simplicity. Axions can be produced in the early universe in a variety of ways, e.g., the misalignment mechanism~\cite{Dine:1982ah,Preskill:1982cy,Turner:1985si,Turner:1983he,Abbott:1982af}, kinetic misalignment~\cite{Co:2019jts,Co:2020dya}, and decays of cosmic strings~\cite{Hagmann:1998me,Gorghetto:2018myk,Battye:1994au,Dine:2020pds,Hindmarsh:2021zkt,Buschmann:2021sdq,Buschmann:2019icd,Klaer:2017qhr,Hiramatsu:2010yu,Gorghetto:2020qws,Hindmarsh:2019csc,Benabou:2024msj}.

The low-energy phenomenology of axions depends on the high-energy theory. The most well-studied of these interactions is the axions coupling to a field strength tensor and its dual, e.g., $\mathcal{L} \supset - g_{a\gamma\gamma} a F^{\mu \nu} \tilde{F}_{\mu\nu} / 4$ where $g_{a\gamma\gamma}$ is the axion-photon coupling, $a$ is the axion, and $F^{\mu \nu}$ is the electromagnetic field strength tensor. This electromagnetic interaction causes an axion in an external magnetic field to convert to an electric field~\cite{Sikivie:1983ip}, and is the operating principle of the many ongoing cavity haloscope experiments; for a recent review see Ref.~\cite{Berlin:2024pzi}. Additionally, axions can couple derivatively to fermionic axial currents~\cite{Berlin:2023ubt}, for example,
\begin{align}
    \mathcal{L} \supset g_{ae} \left( \partial_\mu a \right) \bar{\psi} \gamma^\mu \gamma^5 \psi \, ,
    \label{eq:axion_fermion_interaction}
\end{align}
where $g_{ae}$ is the dimensionful axion-electron coupling, and $\psi$ is the electron. While one generically expects that $g_{a\gamma\gamma}$ and $g_{ae}$ are of the same order, there are axion models where $g_{ae}$ is parametrically larger. For example, if $g_{a\gamma\gamma}$ is zero at high energies the Lagrangian has an enhanced shift symmetry and therefore, in the massless axion limit, $g_{a\gamma\gamma}$ is \textit{not} generated when running the coupling to low energies. These ``photophobic" axions~\cite{Craig:2018kne} only couple via Eq.~\eqref{eq:axion_fermion_interaction}, cannot be detected with traditional cavity haloscope experiments, and highlight the need for new experimental paradigms.

Currently, the most stringent bounds on $g_{ae}$ come from measurements of stellar cooling~\cite{Gondolo:2008dd,MillerBertolami:2014rka,Capozzi:2020}, and XENONnT searches for axions emitted from the Sun~\cite{XENON:2022ltv}. Searching beyond these, into the well-motivated parts of QCD axion parameter space for $1\, \mu\text{eV} \lesssim m_a \lesssim 1\,\text{eV}$, requires dedicated direct detection experiments. Magnons, collective electronic spin excitations, are a particularly promising avenue~\cite{Barbieri:1985cp,Mitridate:2020kly,Chigusa:2020gfs,Catinari:2024ekq} since they can resonate with, or absorb, axions in the mass range of interest. In the simplest magnetic targets, axions are dominantly absorbed into the Kittel mode, whose energy is equal to the Larmor frequency, $\omega_L = 2 \mu_B B_0 \sim 120 \, \mu \text{eV} \, (B_0 / 1 \, \text{T})$, where $\mu_B$ is the Bohr magneton, and $B_0$ is a tunable external magnetic field. This is the operating principle underlying the QUAX experiment~\cite{Ruoso:2015ytk,Crescini:2018qrz,QUAX:2020adt}, which reads out the magnons via coupling to a cavity, and has placed limits on $g_{ae}$ for axions in the $42.4 \, \mu\text{eV} \lesssim m_a \lesssim 43.1 \, \mu \text{eV}$ range. However, these limits are still two orders of magnitude away from the stellar cooling bounds, and the QUAX readout scheme is currently limited by the ``quantum noise'' associated with linear signal amplification. Therefore it is imperative to find alternate readout strategies which can go beyond this ``standard quantum limit (SQL).'' 

Recent work~\cite{Mitridate:2020kly,Chigusa:2020gfs,Catinari:2024ekq} has shown that a detector utilizing single-magnon detection can be orders of magnitude more sensitive to electron-coupled axions. The sensing of individual magnons was recently demonstrated by placing a single-crystal yttrium iron garnet (YIG) sphere and a superconducting transmon qubit into a 3D microwave cavity~\cite{LachanceQuirionScience2020,YouPRL23,RaniarXiv24} and using the superconducting qubit to carry out high-speed quantum non-demolition measurements of the magnon system. Ref.~\cite{Ikeda:2021mlv} analyzed this data to set a limit on $g_{ae}$, and while the exposure of this experiment was too small to yield a competitive axion DM limit, it demonstrated the viability of single-magnon detection to search for axion DM.

\begin{figure}[ht!]
    \centering
    \includegraphics[width=3.0in]{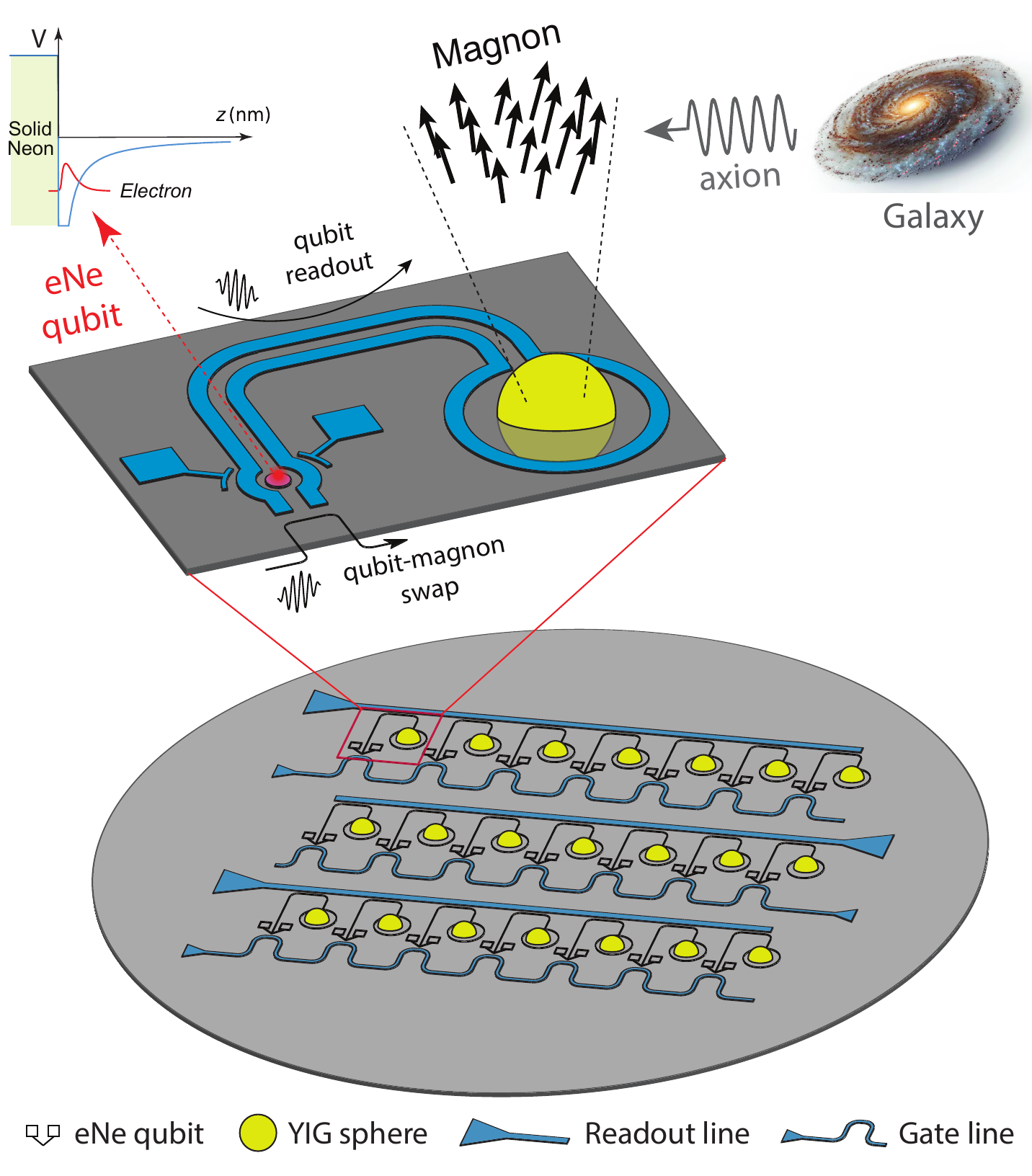}
    \caption{Schematic of the wafer-integrated YIG-qubit detector array (MOSAIC). Each detector tile consists of a coplanar superconducting resonator (blue ring antenna) which couples inductively to a single-crystal YIG sphere (yellow) on one end, and capacitively to an eNe charge qubit (red) on the other.}
    \label{fig:experimental_design}
\end{figure}

In this \textit{Letter}, we present a novel experimental concept using a scalable single-magnon quantum sensor (MOSAIC, Fig.~\ref{fig:experimental_design}) to search for electron-coupled axion DM. The sensing element for MOSAIC combines the recently developed electron-on-solid-neon (eNe) qubit~\cite{zhou2022single,zhou2024electron} with a superconducting coplanar hybrid magnonic circuit~\cite{LiPRL22,SongarXiv2023} to form a chip-embedded magnon-qubit entangled system. The magnetic field tolerance of the eNe qubit, a feature not shared by typical superconducting transmon qubits \cite{SchneiderPRR19}, allows for highly compact integration of the quantum magnon sensor into large, wafer-scale arrays. We find that by using current quantum magnonics technology, a detector array with sensitivity beyond previous ferromagnetic haloscopes can be built and presents a platform where improvements would enable searching for axion DM in unexplored parameter space. In the remainder of this \textit{Letter}, we follow similar discussions of novel axion DM search concepts (e.g., Ref.~\cite{BREAD}) and discuss the expected signal, the sensing concept, estimated sensor performance, and estimate the scientific reach. In our discussion, we consider both a {\it conservative} configuration associated with demonstrated techniques, and an {\it optimistic} configuration where we make reasonable extrapolations of performance. We conclude with a discussion of key challenges illuminated by our analysis and how ongoing developments in quantum sensing have potential to directly improve the scientific reach of our concept.

\vspace{0.5cm}
\noindent
\textbf{Axion Signal.} Before detailing the specific experimental design, we provide a brief overview of the axion-induced signal, which has been explored in depth in Refs.~\cite{Chigusa:2020gfs,Mitridate:2020kly}. In the non-relativistic limit the axion-electron interaction in Eq.~\eqref{eq:axion_fermion_interaction} gives rise to two dominant effects: an ``axion wind" which torques spins, and an ``axioelectric" force~\cite{Berlin:2023ubt}. While both the axion wind and axioelectric force interact with the electron spin, magnon excitations in the $1 \, \mu\text{eV} \lesssim \omegam \lesssim 1 \, \text{meV}$ frequency range are dominantly produced by the axion wind. In the continuum limit, the axion-electron spin interaction Hamiltonian density is $\delta \mathcal{H} = 2 \, g_{ae} \, \nabla a \cdot \vec{s}$, where $\vec{s}(\vec{x})$ is the target electron spin density. When the axion mass, $m_a$, is kinematically matched to $\omegam$ absorption can occur, where an axion converts to a magnon excitation. This corresponds to axion masses well into the ``wave-like" DM limit, where the axion may be treated as a classical field, $a(\vec{x}, t) \approx a_0 \cos( m_a t - \vec{p}_a \cdot \vec{x})$, $\rho_a = m_a^2 a_0^2 / 2 \approx 0.4 \, \text{GeV} / \text{cm}^3$ is the local axion DM density, $\vec{p}_a \approx m_a \vec{v}_a$ is the axion momentum, and $v_a \sim 10^{-3}$ is the axion velocity; these quantities are collectively informed by particle-astrophysical models~\cite{Cirelli:2024ssz}. Furthermore, the gradient of the axion field may be taken to be homogeneous since its wavelength, $\lambda_a = 2 \pi / (m_a v_a) \sim 1\, \text{m} \, \left( \text{meV} / m_a \right)$, is much larger than our experimental apparatus. Therefore the axion-electron interaction Hamiltonian is simply $\delta H \approx 2 m_a a_0 g_{ae} \cos{(m_a t)} \, \vec{v}_a \cdot \vec{S}$, where $\vec{S}$ is the total spin of the target, summed over all individual spins,  which is quantized in terms of the magnon degrees of freedom~\cite{Toth_2015}. 

This interaction Hamiltonian dominantly causes transitions between the ground state and a single Kittel mode magnon state. The absorption rate per target mass into the Kittel mode can be calculated with Fermi's Golden rule as~\cite{Chigusa:2020gfs,Mitridate:2020kly},
\begin{align}
    R \approx g_{ae}^2 \frac{\rho_a}{\rho_T} n_s v_a^2 \sin^2{\theta} \frac{4 \, m_a \, \omega_\text{m} \, \gamma_\text{m}}{(m_a^2 - \omega_\text{m}^2)^2 + ( m_a \gamma_\text{m} )^2} \, ,
    \label{eq:R_abs}
\end{align}
where $\rho_T$ is the target mass density, $n_s$ is the target spin density, $\cos{\theta} = \hat{\vec{z}} \cdot \hat{\vec{v}}_a$, where $\hat{\vec{z}}$ is defined to be parallel to the magnetization, and the Lorentzian appears from broadening the delta function in Fermi's Golden rule with a magnon linewidth $\gamma_\text{m}$. On resonance, $m_a = \omegam$, this rate is given parametrically as,
\begin{align}
    R_\text{res} \sim \frac{1}{\text{mg} \cdot \text{yr}} \left( \frac{g_{ae}}{10^{-10} \, \text{GeV}^{-1}} \right)^2 \left( \frac{1 \,  \text{MHz}}{\gamma_\text{m} / 2 \pi} \right) \, ,
    \label{eq:R_abs_parametric}
\end{align}
where we average over the incoming axion velocity such that $\sin{\theta} \rightarrow \sqrt{2 / 3}$, and assume YIG material parameters: $n_s \approx 2 \times 10^{22} / \text{cm}^{3}$, $\rho_T \approx 5.1 \, \text{g} / \text{cm}^3$. The magnon linewidth, $\gamma_\text{m}$, has been measured to be a frequency-independent value of $\gamma_\text{m} / 2 \pi = 1 \, \mathrm{MHz}$ at 10 mK temperatures~\cite{TabuchiPRL14}, which holds up to 40 GHz where the Gilbert damping linewidth broadening starts to dominate~\cite{KlinglerAPL17}. The total rate of magnon excitation for a detector of mass $m_\text{YIG}$ is $\Gamma = R \times m_\text{YIG}$.

\vspace{0.5cm}
\noindent
\textbf{Detector Design And Operation.} The MOSAIC detector (see Fig.~\ref{fig:experimental_design}) is an array of many individual sensing elements (\textit{tiles}), integrated onto wafers, and housed within a commercial dilution refrigerator operating at a temperature of $T = 10 \, \text{mK}$ and subjected to a tunable external magnetic field of strength $B_0$. Each tile features a coplanar superconducting resonator (CSR) with frequency $\omega_\text{r}$, which is inductively coupled to the Kittel mode of a single-crystal YIG sphere at its $H$-field maximum, and capacitively coupled to an eNe qubit at its $E$-field maximum. The magnon frequency is determined by $B_0$, and the eNe charge qubit frequency, $\omega_\text{q}$, is determined by the gating electrodes. Varying both of these external controls allows the sensor to be tuned to different frequencies. 

The size of the YIG sphere is crucial for determining the sensitivity of the detector. Large YIG masses are obviously preferred, and, in principle, it is possible to incorporate large YIG masses with complex resonator designs, where the size of the YIG mass is limited by the efficiency of the resonator-YIG coupling. However, these techniques and their limitations have not been well-developed. So, for the {\it conservative} configuration of MOSAIC, we take $d_\text{YIG} = 1 \, \text{mm}$ ($m_\text{YIG} = 2.7 \, \text{mg}$), since this size has been used in the demonstrated single-magnon quantum sensing \cite{YouPRL23,RaniarXiv24}, and can be also easily embedded in a coplanar superconducting circuit~\cite{LiPRL22}. For our \textit{optimistic} configuration, we assume $d_\text{YIG} = 10 \, \text{mm}$ ($m_\text{YIG} = 2.7 \, \text{g}$), as this has a linear dimension that is a fraction of the resonator wavelength at several GHz. Additionally, we expect an in-situ $\omega_\text{q}$ tunability of $\Delta \omega_\text{q} / 2 \pi = 1 \, \text{GHz}$ based on current measurements~\cite{Zhou:2021ylr,Li:2025mxv}.

To count the magnons generated in each YIG sphere we dispersively couple the magnon to the qubit via the CSR. The effective magnon-qubit interaction Hamiltonian is~\cite{Koch07}:
\begin{align}
    \hat{H}_\text{q-m} = \left( \omega_\text{q} + 2\, \chi_\text{q-m} \, \hat{b}^\dagger\hat{b} \right) \, \hat{a}^\dagger \hat{a} \, ,
    \label{eq:Ham}
\end{align}
where $\hat{a}^\dagger$ ($\hat{a}$) is the qubit creation (annihilation) operator, $\hat{b}^\dagger$ ($\hat{b}$) is the magnon creation (annihilation) operator and $\chi_\text{q-m}$ is the dispersive qubit frequency shift. From Eq.~\eqref{eq:Ham}, we see that the magnon number-shifted qubit frequency, $\omega_\text{q}^N$, depends on the magnon occupation number, $N_\text{m}=\langle \hat{b}^\dagger\hat{b} \rangle$, as $\omega_\text{q}^N=\omega_\text{q}+2\,\chi_\text{q-m} \, N_\text{m}$. Detecting this qubit dispersive shift, and thus the presence/absence of a single magnon, is achieved through the following three step quantum non-demolition measurement protocol \cite{LachanceQuirionScience2020}:
\begin{itemize}
    \item \textit{Step 1. Qubit-magnon swap}: Apply a microwave Rabi pulse at $\omega_\text{q}$ through the gate line and excite the qubit conditionally when $N_\text{m}=0$ (no excitation in the presence of one or more magnons, $N_\text{m}>0$).
    \item \textit{Step 2. Qubit readout}: Apply a microwave pulse at $\omega_\text{r}$ through the readout line and measure the CSR's dispersive frequency shift to access the qubit state.
    \item \textit{Step 3. Qubit reset}: Apply a qubit gate pulse through the gate line that resets the qubit state to its ground state.
\end{itemize}
This protocol has been successfully demonstrated in magnon quantum sensing with a 3D cavity and a transmon qubit~\cite{LachanceQuirionScience2020}.

The total detector cycle time $\Delta\tau$ for the above protocol is also an important factor in detector sensitivity. Ideally, the entire protocol would be performed once every magnon lifetime, $\tau_\text{m} = 1 / \gamma_\text{m} = 160 \, \text{ns}$, allowing the detector to sense any generated magnon. In practice, steps 1-3 will each take time, $\tau_i$ for $i = 1, 2, 3$, respectively. For step 1, the minimal duration of the Rabi pulse is set by the bandwidth of the dispersive qubit-magnon frequency shift $\chi_\text{q-m}$. Targeting a $\chi_\text{q-m}/2\pi=5$ MHz, for scalability reasons that will be explained below, yields $\tau_1=2\pi/2\chi_\text{q-m}=100$ ns. Steps 2 and 3 can potentially be conducted within 100 ns, which we take as $\tau_2$ and $\tau_3$, using a quantum-limit parametric amplifier for qubit readout (step 2)~\cite{DassonnevillePRX20,SunadaPRXQ24}, and an active pulse reset protocol for qubit reset (step 3)~\cite{EggerPRApplied18,SunadaPRApplied22}. Therefore the total time for each measurement is $\Delta \tau = \tau_1 + \tau_2 + \tau_3 = 300 \, \text{ns}$.

Another important factor in determining the detector sensitivity is the dark count probability, which is primarily due to the errors of qubit operations. eNe qubits have a demonstrated single-shot fidelity of $98.1\%$~\cite{zhou2024electron}, corresponding to a dark count probability of $\bar{p}_\text{dc} = 1.9 \times 10^{-2}$, which we assume for the {\it conservative} MOSAIC configuration. This value is limited by the fidelity of step 2. For the {\it optimistic} configuration, we use the eNe qubit gate fidelity from steps 1 and 3 (measured to be a much higher 99.97\%~\cite{zhou2024electron}) as a lower limit, which gives a value of $\bar{p}_\text{dc} = 3 \times 10^{-4}$.

Lastly, we discuss the ability to multiplex many individual sensing tiles, which is a unique advantage of the MOSAIC detector since we do not rely on a 3D cavity. Fig. ~\ref{fig:experimental_design} shows a planar schematic where we couple multiple detectors to a single coplanar waveguide (CPW) feedline for multiplexing readout. The multiplexing bandwidth is the maximal range of frequency detuning between the eNe charge qubit and magnon system, $\Delta_\text{q-m}=\omega_\text{q}-\omega_\text{m}$, for which the system can still fulfill high-speed and high-fidelity single-magnon detection. Achieving this requires large $\chi_\text{q-m}$, related to $\Delta_\text{q-m}$ by~\cite{Koch07,BlaisRMP21}:
\begin{align}
    \chi_\text{q-m}=\frac{\alpha \, g^2_\text{q-m}}{\Delta_\text{q-m}(\Delta_\text{q-m}+\alpha)} \, ,
    \label{eq:dispersive}
\end{align}
where $\alpha /2 \pi = 1 \, \text{GHz}$ ~\cite{zhou2024electron} is the anharmonicity of the eNe charge qubit and $g_\text{q-m}$ is the magnon-qubit coupling strength mediated by the CSR. Since the magnon-qubit interaction is mediated by the cavity photon and the 2D-CSR-based magnon-cavity coupling ~\cite{LiPRL19,LiPRL22} is much higher than the qubit-cavity coupling, the main limitation of $g_\text{q-m}$ for an eNe qubit is the qubit-cavity coupling, with an anticipated improvement of up to $g_\text{q-m} / 2 \pi = 43 \text{MHz}$ with high-impedance CSR design~\cite{KoolstraPRApplied25}. From Eq.~\eqref{eq:dispersive} we find that reaching $\chi_\text{q-m}/2\pi \ge 5 \,\text{MHz}$ and keeping $\tau_1 \le 100 \, \text{ns}$, is satisfied for $|\Delta_\text{q-m}|/2\pi \le 185 \, \text{MHz}$. Assuming that $\omega_\text{q}$ for different qubits must be well separated by, e.g., $5\cdot2\chi_\text{q-m} / 2 \pi  = 50 \, \text{MHz}$ to avoid cross-talk sets the multiplexing number of qubits per CPW line to 7, with only 2 coaxial lines needed. This shows the capability of detector multiplexing and being able to integrate more than one detector per coax line. With state-of-the-art dilution refrigerators hosting over 1000 coax lines~\cite{Bluefors}, $N_s = 3500$ individual tiles can be operated following the architecture in Fig.~\ref{fig:experimental_design}. For even larger $N_s$, the main limitation in increasing $N_s$ comes from circuit-level scaling challenges~\cite{Martinis24}, and we assume $N_s = 10^4$ for the {\it optimistic} MOSAIC configuration.

\begin{figure}[ht!]
    \centering
    \includegraphics[width=\linewidth]{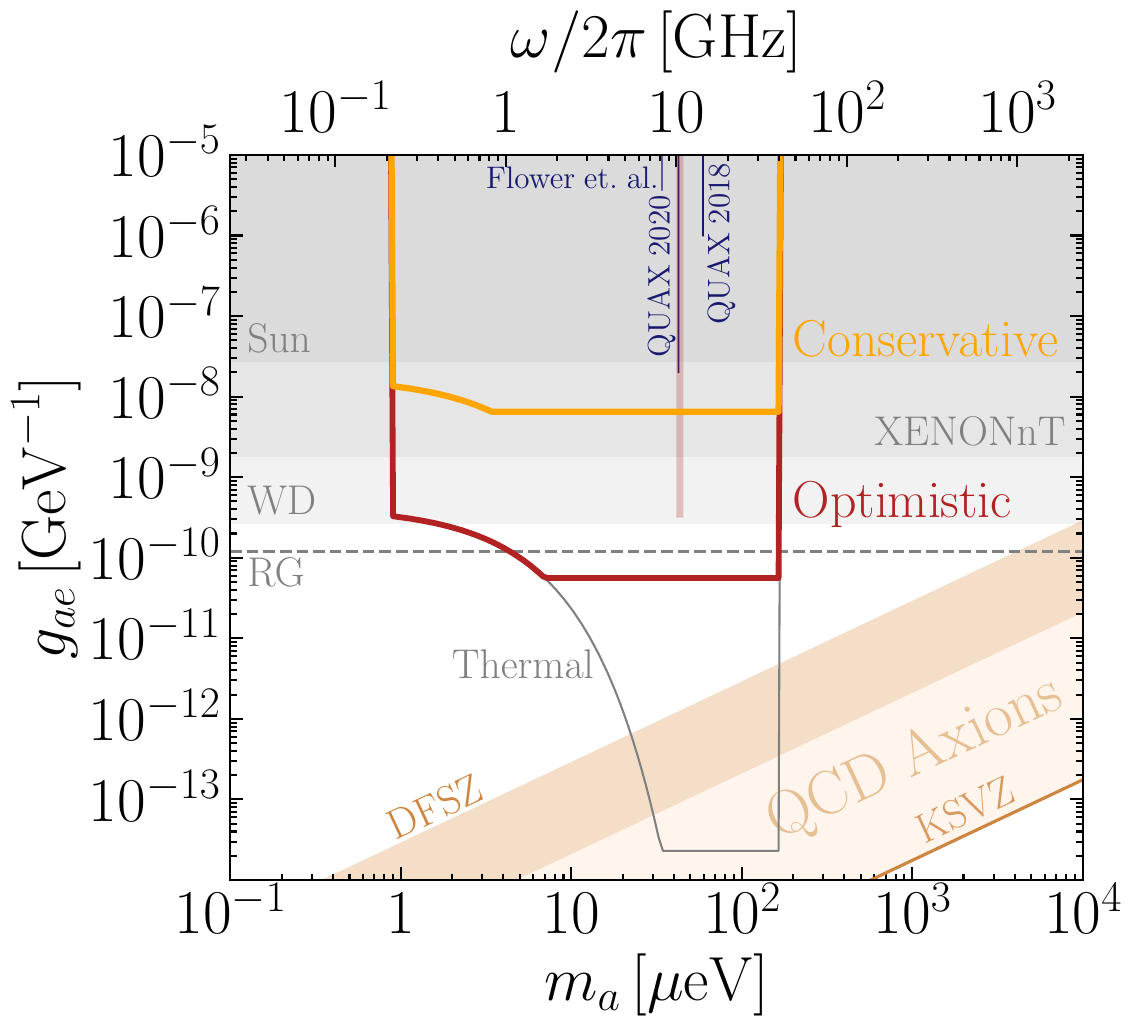}
    \caption{Projected sensitivity to the axion-electron coupling $g_{ae}$, Eq.~\eqref{eq:axion_fermion_interaction}, with the MOSAIC detector (Fig.~\ref{fig:experimental_design}) operating at a temperature of $T = 10 \, \text{mK}$ for $T_\text{obs} = 1 \, \text{yr}$. The orange line assumes the \textit{conservative} detector parameters, with component performance that has been demonstrated in the quantum magnonics community, and the red line assumes a setup with the \textit{optimistic} detector parameters. The shaded red region is the scanning sensitivity with the \textit{optimistic} detector parameters; with one year of observation the entire shaded region can be excluded. The gray line (``Thermal") is the projected sensitivity if only thermal backgrounds contribute. Shaded gray regions are bounds from stellar cooling in the Sun~\cite{Gondolo:2008dd} and white dwarfs (WD)~\cite{MillerBertolami:2014rka}, and from the XENONnT experiment~\cite{XENON:2022ltv}. The dashed gray line is from cooling of red giant (RG) stars~\cite{Capozzi:2020}, however the robustness of these bounds has been questioned~\cite{Dennis:2023kfe}. Existing bounds from QUAX~\cite{Crescini:2018qrz,QUAX:2020adt} and Ref.~\cite{Ikeda:2021mlv} (``Flower et. al.") are shown in blue. The gold shaded region (``QCD Axions") is the part of parameter space where the axion can also solve the strong CP problem~\cite{OHare:2024nmr}, and dark gold regions correspond to the specific DFSZ and KSVZ models~\cite{Berlin:2023ubt}.}
    \label{fig:projected_sensitivity}
\end{figure}

\vspace{0.5cm}
\noindent
\textbf{Detector Sensitivity.} The MOSAIC detector has many attractive properties including single-magnon sensitivity of a substantial target volume across a range of frequencies. We now consider the sensitivity of the MOSAIC detector shown in Fig.~\ref{fig:experimental_design}, to the axion-induced signal discussed before Eq.~\eqref{eq:R_abs}. We limit our analysis to axion masses (frequencies) in the range $200 \, \text{MHz} \lesssim m_a / 2 \pi \lesssim 40 \, \text{GHz}$; the smaller frequency corresponds to when the thermal magnon population is significant ($m_a \sim T$), and the larger frequency corresponds to the operating limit of the microwave electronics. We assume an array of $N_s$ magnon-qubit detectors running in parallel over an observation time of $T_\text{obs}$, with a detector cycle time $\Delta \tau$, and a magnon creation probability of $p_\text{m}$. If only one mode at frequency $\omega_\text{m} = m_a$ is interrogated, then over the observation time the number of observations is $N_\text{obs} = T_\text{obs} / \Delta \tau$. The expected number of signal magnons is then $N_\text{sig} =  N_s \, N_\text{obs} \times p_\text{m} = N_s \, N_\text{obs} \times( R \, m_\text{YIG} \, \tau_\text{m} )$, where $R$ is the magnon excitation rate in Eq.~\eqref{eq:R_abs}. Magnon and qubit frequency tunability provides an ideal method for removing systematic uncertainties. By comparing the number of events in modes nearby in frequency, backgrounds can be modeled. In the limit backgrounds can be understood, the signal to noise ratio, SNR, is given statistically as, $\text{SNR} = N_\text{sig} / \sqrt{N_\text{dc}}$, where $N_\text{dc}$ is the number of dark counts~\cite{Chen:2022quj,Linehan:2024btp}.

The number of dark, or background, counts is $N_\text{dc} = N_s N_\text{obs} \times p_\text{dc}$, where $p_\text{dc}$ is the probability of a dark count from a single MOSAIC tile. We expect the two main sources of dark counts are from a thermal population of magnons, and qubit readout error. Therefore we parameterize the dark count probability as, $p_\text{dc} = \text{max}\{ \bar{p}_\text{dc},\, \exp{(- m_a / T)} \}$. With this parameterization of the dark counts we can understand the sensitivity parametrically to $g_{ae}$ as:
\begin{align}
    g_{ae} \sim \frac{10^{-8}}{\text{GeV}} \, N_s^{- \frac{1}{4} }\left( \frac{2.7 \, \text{mg}}{m_\text{YIG}} \right)^\frac{1}{2} \left( \frac{p_\text{dc}}{3 \times 10^{-4}} \right)^\frac{1}{4} \left( \frac{\tau_\text{m}}{160 \, \text{ns}} \right) \, ,
    \label{eq:gae_parametrics}
\end{align}
where we we have assumed $T_\text{obs} = 1 \, \text{yr}$, and normalized $p_\text{dc}$ to the optimistic value of $\bar{p}_\text{dc}$. The projected sensitivity to the axion-electron coupling versus axion mass is shown in Fig.~\ref{fig:projected_sensitivity} for the {\it conservative} configuration (orange line, labeled ``Conservative"), and {\it optimistic} configuration i.e., heavier YIG spheres, lower dark count probabilities, and further multiplexing (red line, labeled ``Optimistic"). The gray line, labeled ``Thermal", shows the sensitivity of the optimistic detector in the limit only thermal backgrounds contribute ($\bar{p}_\text{dc} = 0$). For $m_a \gtrsim 30 \, \mu\text{eV}$ the number of thermal magnons becomes negligible and the detector can operate in the background-free limit. In the background-free limit we derive the sensitivity by requiring that $N_\text{sig} > 3$, which corresponds to the 95\% C.L. exclusion region if no events are observed. Note that these lines show the sensitivity that can be achieved at any given $m_a$ if the total observation time is spent at $\omega_\text{m} = m_a$.

To cover a range of axion masses the observation time must be split between each signal linewidth. Since the magnon linewidth is larger than the axion linewidth, $\gamma_a \sim m_a / Q_a \sim \text{kHz} \, (m_a / \mu\text{eV})$, where $Q_a \sim 10^6$ is the axion quality factor~\cite{Berlin:2024pzi}, the number of scan steps is determined by $\gamma_\text{m}$. While the magnon frequency is easily tunable with the external magnetic field, the range of axion masses which can be scanned is limited by the in-situ dynamic range of the qubit, $\Delta \omega_\text{q}$. To search for all axion masses in the range $\Delta \omega_\text{q}$, $\Delta \omega_\text{q} / \gamma_\text{m} = 10^3$ scan steps must be performed. This limits the observation time at each step to $t_\text{obs} \approx T_\text{obs} \gamma_\text{m} / \Delta \omega_\text{q}$, and reduces the sensitivity relative to the solid lines in Fig.~\ref{fig:projected_sensitivity}. To illustrate the scanning sensitivity (the shaded red region in Fig.~\ref{fig:projected_sensitivity}) we focus on the $10 \, \text{GHz} \lesssim \omega / 2\pi \lesssim 11 \, \text{GHz}$ frequency range. In one year of observation time \textit{all} of the parameter space in the shaded region can be excluded.

Lastly, we compare the sensitivity of an individual MOSAIC tile to the ``standard quantum limit (SQL)'' limited, linear amplification based detector used by the QUAX collaboration~\cite{Crescini:2018qrz,QUAX:2020adt}. The SNR for readout by linear amplification in a system at noise temperature $T_n$ is given by the Dicke radiometer equation~\cite{Dicke:1946glx}, $\text{SNR} = (P_s / T_n) \sqrt{T_\text{obs} / \Delta \nu}$, where $P_s = m_a R \, m_\text{YIG}$ is the signal power and $\Delta \nu = \omega / (2 \pi Q_a)$ is the bandwidth. At the SQL, $T_n$ is limited to $m_a$~\cite{Caves:1982zz}, and therefore the SNR is limited to $\text{SNR}_\text{SQL} = m_\text{YIG} \, R \,\sqrt{2 \pi Q_a T_\text{obs} / m_a }$. Comparing this SQL limited SNR to the one derived previously we find,
\begin{align}
    \frac{\text{SNR}}{\text{SNR}_\text{SQL}} & \sim \left( \frac{10^{-4}}{p_\text{dc}} \right)^\frac{1}{2} \left( \frac{m_a}{10 \, \mu\text{eV}} \right)^\frac{1}{2} \, ,
    \label{eq:SNR_compare}
\end{align}
for $\tau_\text{m} = 160 \, \text{ns}$ and $\Delta \tau = 300 \, \text{ns}$. Therefore for $m_a \gtrsim 10 \, \mu\text{eV}$, if $p_\text{dc} \lesssim 10^{-4}$ can be achieved the sensitivity of each MOSAIC tile can surpasses the SQL.

\vspace{0.5cm}
\noindent
\textbf{Conclusions.} This \textit{Letter} presents the concept for a magnonic quantum sensor array (MOSAIC, Fig.~\ref{fig:experimental_design}) which can be scaled to exposures necessary to search new parameter space for electron spin-coupled axion DM. This approach complements broader axion searches via the photon-coupling, and builds upon demonstrated techniques in quantum magnonics. The MOSAIC detector is composed of many tiles, each coupling a YIG sphere (magnon system) to an eNe qubit via a coplanar superconducting resonator. Quantum non-demolition measurements are used to measure the magnon population produced in each tile. The magnetic field resilience of the eNe charge qubit permits compact wafer integration of individual tiles, and allows the system to be scaled. We have shown that a MOSAIC detector built in the {\it conservative} configuration with currently available quantum magnonics components could achieve superior sensitivity compared to existing ferromagnetic haloscopes, and an {\it optimistic} configuration could probe new parameter space below stellar cooling bounds. Furthermore the forecasted background-free sensitivity for the {\it optimistic} configuration could test the theoretically well-motivated QCD axion models.

Our analysis highlights the opportunities for improvement and fundamental limitations of this array-scalable, single-magnon counting detector design. For $m_a \lesssim 30 \, \mu\text{eV}$ the sensitivity is limited by the thermal magnon background when operating at $T = 10 \, \text{mK}$. For $30 \, \mu\text{eV} \lesssim m_a$ the achievable sensitivity could be improved in a variety of ways. For example, if multiple YIG spheres can be coherently read out with a single qubit, this would effectively increase $m_\text{YIG}$, which is more favorable than just increasing the numbers of tiles (Eq.~\eqref{eq:gae_parametrics}). Additionally, improvements to the dark count probability, $p_\text{dc}$ (or qubit gate fidelity) would also improve the sensitivity to electron-coupled axion DM. While current magnon-based devices are limited to $p_\text{dc} \gtrsim 3 \times 10^{-4}$, dark count probabilities of $p_\text{dc} \sim 10^{-6}$ have been demonstrated in pumped qubit photon sensors using 4-wave mixing~\cite{BalemboisPRApplied24}. If such an approach could be used here, each tile could achieve sensitivity beyond the SQL. Other quantum sensing techniques offer further possibilities for reducing the dark count rate (improving the qubit gate fidelity). For example, it has been demonstrated that using multiple entangled qubits can improve readout fidelity~\cite{giovannetti2004quantum}. This principle indicates that by coupling multiple entangled qubits to each YIG sphere within our system, we may achieve a reduction in $p_\text{dc}$. Another potential direction would be to improve gate fidelity by incorporating quantum error correction~\cite{zhou2018achieving,zhou2024achieving,zhou2024limits}. The MOSAIC detector array provides a platform for implementing these combined aforementioned improvements to realize a detector potentially sensitive not only to QCD axion DM, but also to the broader class of light DM scattering signals that generate single magnons~\cite{Trickle:2019ovy,Esposito:2022bnu,Marocco:2025eqw,Berlin:2025uka}.

\begin{acknowledgments}
    We would like to thank Dany Lachance-Quirion for discussion on the error rate calculation, and Asher Berlin and Roni Harnik for helpful discussion. This work was supported by Quantum Information Science Enabled Discovery 2.0 (QuantISED 2.0) for High Energy Physics (DE-FOA-0003354), with work at Argonne National Laboratory supported by the U.S.~Department of Energy under contract DE-AC02-06CH11357.  This manuscript has been authored by Fermi Research Alliance, LLC under Contract No. DE-AC02-07CH11359 with the U.S. Department of Energy, Office of Science, Office of High Energy Physics.  Y.L. and V.N. acknowledge support by the U.S. Department of Energy, Office of Science, Basic Energy Sciences, Materials Sciences and Engineering Division under Contract No. DE-SC0022060. D.J. acknowledges support by the Department of Energy (DOE) under Award No. DE-SC0025542 for research effort on solid neon and the Air Force Office of Scientific Research (AFOSR) under Award No. FA9550-23-1-0636 for research effort on charge qubits. 
\end{acknowledgments}

\bibliographystyle{utphys3}
\bibliography{bibliography}

\end{document}